\documentclass[st,twocolumn]{jpsj3}
\usepackage{txfonts}
  
\usepackage{bm}
\usepackage{amsmath}
\usepackage{mathtools}
\usepackage{wrapfig}

\newcommand*{\fleur}{\texttt{FLEUR}}
\newcommand*{\wprog}{\textsc{wannier}{\footnotesize{90}}}
\newcommand*{\DM}{Dzyaloshinskii-Moriya}
\newcommand*{\vn}[1]{\bm{\mathrm {#1}}}

\newcommand*{\gwfshort}{HDWFs}
\newcommand*{\gpos}{\vn \xi}
\newcommand*{\gdlatt}{\vn \Xi}
\newcommand*{\gpara}{\vn \lambda}

\newcommand*{\gwann}{W}
\newcommand*{\gumat}{\mathcal{U}}
\newcommand*{\artpsi}{\zeta}
\newcommand*{\artu}{\rho}

\title{Higher-dimensional Wannier interpolation for the modern theory of the Dzyaloshinskii-Moriya interaction: Application to Co-based trilayers}

\author{Jan-Philipp Hanke\thanks{j.hanke@fz-juelich.de}, Frank Freimuth, Stefan Bl\"ugel, and Yuriy Mokrousov}
\inst{Peter Gr\"unberg Institut and Institute for Advanced Simulation,\\Forschungszentrum J\"ulich and JARA, 52425 J\"ulich, Germany} 

\abst{We present an advanced first-principles formalism to evaluate the Dzyaloshinskii-Moriya interaction (DMI) in its modern theory as well as Berry curvatures in complex spaces based on a higher-dimensional Wannier interpolation. Our method is applied to the Co-based trilayer systems Ir$_\delta$Pt$_{1-\delta}$/Co/Pt and Au$_\gamma$Pt$_{1-\gamma}$/Co/Pt, where we gain insights into the correlations between the electronic structure and the DMI, and we uncover prominent sign changes of the chiral interaction with the overlayer composition. Beyond the discussed phenomena, the scope of applications of our Wannier-based scheme is particularly broad as it is ideally suited to study efficiently the Hamiltonian evolution under the slow variation of very general parameters.}

\begin{document}
\maketitle

\section{Introduction}

Embedding the basic ideas of spintronics into future information technologies relies crucially on our ability to control and manipulate efficiently the magnetic state of nanomagnets, for example, by utilizing relativistic effects. In this context, the so-called \DM{} interaction (DMI)~\cite{Dzyaloshinsky1958,Moriya1960} experiences revived prominence as it can stabilize chiral magnetic textures in inversion-asymmetric crystals with spin-orbit coupling.~\cite{Bode2007} The DMI facilitates in particular the formation of topologically non-trivial magnetic skyrmions,~\cite{Bogdanov1989,Neubauer2009,Kanazawa2011,Freimuth2013a} and assists the current-induced ultrafast motion of domain walls.~\cite{Tretiakov2010,Thiaville2012,Haazen2013,Ryu2013,Emori2013} Recently, the magnitude of the interfacial DMI was shown to be tunable in multilayers of Co in between heavy metals such as Pt and Ir,~\cite{Hrabec2014,MoreauLuchaire2016,Yang2015a} which holds bright prospects for the observation of small skyrmions at room temperature.

A commonly pursued approach to predict from {\itshape ab initio} the magnitude of the DMI is to perform non-collinear calculations of spin spirals with wave vector $\vn q$, where spin-orbit coupling is added as a small perturbation.~\cite{Heide2009,Zimmermann2014,Schweflinghaus2016} Then, the chiral interaction manifests in the $\vn q$-linear part of the dispersion $E(\vn q)$. Alternatively, the DMI can be extracted from derivatives of the spin correlation function,~\cite{Koretsune2015} intrinsic spin currents,~\cite{Kikuchi2016} a tight-binding band representation,~\cite{Katsnelson2010} or from multiple-scattering theory.~\cite{Mankovsky2017} Some of these approaches require notably heavy computational frameworks associated with non-collinear spin structures or the full-relativistic Kohn-Korringa-Rostoker method. Others rely on limiting cases for exchange and spin-orbit interactions, rendering the accuracy of the results for DMI in materials that include heavy metals such as the aforementioned multilayers questionable. In contrast, a Berry phase theory was developed in Ref.~\citen{Freimuth2014} that evaluates the DMI solely from the electronic structure of the collinear ferromagnetic ground state with self-consistent spin-orbit coupling. This Berry phase formalism is also referred to as modern theory of DMI.

Here, we promote an advanced first-principles scheme that is ideally suited to describe efficiently the DMI in the modern theory based on the ferromagnetic state with magnetization direction $\hat{\vn m}$. Our method facilitates the accurate prediction of DMI and the intimately related spin-orbit torques for a wide community of researchers that work routinely with the Wannier interpolation. As an example, we apply our method to the Co-based trilayers Ir$_\delta$Pt$_{1-\delta}$/Co/Pt and Au$_\gamma$Pt$_{1-\gamma}$/Co/Pt, where we uncover an intricate interplay between the overlayer composition and the chiral interaction. By emphasizing the correlations with the electronic structure of these systems, we analyze the microscopic origin of the DMI and identify its anisotropy with the magnetization direction.

\section{Theory}
\paragraph{Modern theory of DMI}The DMI can be quantified by the so-called spiralization tensor $D(\hat{\vn m})$ reflecting the change of the micromagnetic free energy density $F(\vn r)$ due to chiral perturbations $\partial\hat{\vn m}(\vn r)/\partial r_j$, which assumes up to first order in magnetization gradients the form~\cite{Freimuth2014}
\begin{equation}
F^{(1)}(\vn r)=\sum_{ij}D_{ij}(\hat{\vn m})\,\hat{\vn e}_i\cdot\left(\hat{\vn m}(\vn r)\times \frac{\partial\hat{\vn m}(\vn r)}{\partial r_j}\right) \, ,
\label{eq:free_energy_change}
\end{equation}
where $\hat{\vn e}_i$ is the $i$th Cartesian unit vector, and $\vn r$ is the position. We focus in this work on the DMI in ferromagnetic systems, for which the magnetization direction $\hat{\vn m}$ exhibits only a weak dependence on the position $\vn r$. Thus, an expression for the spiralization tensor can be obtained if the free energy change due to small spatial oscillations of the magnetization direction is equated with the free energy change according to Eq.~\eqref{eq:free_energy_change}.

We define the torque operator $\vn{\mathcal T}= \hat{\vn m}\times \vn B_\mathrm{xc}$ based on the exchange field $\vn B_\mathrm{xc}$, and the velocity operator in crystal momentum representation, $\hbar\vn v(\vn k) = \partial_{\vn k} (e^{-i\vn k \cdot \vn r} \hat H e^{i \vn k \cdot \vn r})$, where $\hat H$ is the single-particle Hamiltonian of the ferromagnetic system. Following the detailed derivation in Ref.~\citen{Freimuth2014}, we find that the spiralization at finite temperatures $T$ amounts to
\begin{equation}
D_{ij} = \frac{1}{NV} \sum_{\vn k n}\bigg[f(\mathcal E_{\vn k n}) A_{ij}^n(\vn k) + \frac{1}{\beta} \ln \left(1+e^{-\beta (\mathcal E_{\vn kn}-\mu)}\right) B_{ij}^n(\vn k)\bigg] \, ,
\label{eq:spiralization_finiteT}
\end{equation}
where $N$ is the number of $\vn k$-points, $V$ is the unit cell volume, $f(\mathcal E_{\vn kn})$ is the Fermi distribution function with the band energy $\mathcal E_{\vn kn}$, $\beta=1/(k_\mathrm{B}T)$, and $\mu$ is the chemical potential. The intrinsic scattering-independent nature of Eq.~\eqref{eq:spiralization_finiteT} is mediated by the quantities $A_{ij}^n$ and $B_{ij}^n$ hinging on matrix elements of the torque and velocity operators:
\begin{equation}
A_{ij}^n(\vn k) = -\mathrm{Im}\sum_{m\neq n}\frac{\langle u_{\vn kn} | \mathcal T_i | u_{\vn k m}\rangle \langle u_{\vn k m} | \hbar v_j(\vn k) | u_{\vn k n}\rangle}{\mathcal E_{\vn k n}-\mathcal E_{\vn k m}} \, ,
\label{eq:spiralization_A}
\end{equation}
and
\begin{equation}
B_{ij}^n(\vn k) = -2\mathrm{Im}\sum_{m\neq n}\frac{\langle u_{\vn kn} | \mathcal T_i | u_{\vn k m}\rangle \langle u_{\vn k m} | \hbar v_j(\vn k) | u_{\vn k n}\rangle}{(\mathcal E_{\vn k n}-\mathcal E_{\vn k m})^2} \, .
\label{eq:spiralization_B}
\end{equation}
Here, the $|u_{\vn k n}\rangle$ denote the lattice-periodic parts of the Bloch wave functions $|\psi_{\vn kn}\rangle=e^{i\vn k \cdot \vn r} | u_{\vn kn}\rangle$. In order to elucidate the deep geometric origin of Eq.~\eqref{eq:spiralization_finiteT}, we rewrite the torque operator as gradient of the Hamiltonian with respect to $\hat{\vn m}$, i.e., $\vn{\mathcal T}=\hat{\vn m} \times \partial_{\hat{\vn m}} \hat H$, and represent the magnetization direction as $\hat{\vn m}=(\cos\varphi\sin\theta,\sin\varphi\sin\theta,\cos\theta)^\mathrm{T}$ using the azimuthal angle $\varphi$ and the polar angle $\theta$ shown in Fig.~\ref{fig:fig1}(b). Then, an alternative expression for the spiralization at zero temperature is obtained in terms of derivatives of the wave functions with respect to the crystal momentum and the magnetization direction:
\begin{equation}
\begin{split}
D_{ij} &= \frac{1}{NV} \sum_{\vn k n}^{\mathrm{occ}} \left[ A_{ij}^n(\vn k) - (\mathcal E_{\vn kn}-\mathcal E_\mathrm{F}) B_{ij}^n(\vn k)\right] \\
&=\frac{\hat{\vn e}_i}{NV}\cdot \mathrm{Im}\sum_{\vn k n}^{\mathrm{occ}} \bigg[\hat{\vn m} \times \bigg\langle \frac{\partial u_{\vn kn}}{\partial \hat{\vn m}} \bigg| h_{\vn kn} \bigg| \frac{\partial u_{\vn kn}}{\partial k_j}\bigg\rangle \bigg] \, ,
\end{split}
\label{eq:spiralization_zeroT}
\end{equation}
where $h_{\vn kn}=H_{\vn k} + \mathcal E_{\vn kn} - 2\mathcal E_\mathrm{F}$ with $H_{\vn k}=e^{-i\vn k \cdot \vn r}\hat H e^{i\vn k \cdot \vn r}$, the sum is restricted to all occupied states, and $\mu$ was replaced with the Fermi level $\mathcal E_\mathrm{F}$. The Berry phase expression~\eqref{eq:spiralization_zeroT} of the spiralization is strongly reminiscent of the modern theory of orbital magnetization~\cite{Thonhauser2011,Resta2010,Xiao2010}. It is therefore tempting to refer to Eq.~\eqref{eq:spiralization_zeroT} as the modern theory of DMI, which has a manifestly geometric contribution originating from the adiabatic Hamiltonian evolution of the ferromagnetic system under slow variations of the crystal momentum and the magnetization direction.

To model the effect of disorder, the Kubo linear-response theory of the DMI spiralization in the clean limit\cite{Freimuth2014} can be extended to include a constant band broadening $\Gamma$, which yields
\begin{align}
\nonumber D_{ij} &= \frac{1}{2\pi N V} \sum_{\vn k n}\sum_{m\neq n} \mathrm{Im}\left[ \langle u_{\vn kn} | \mathcal T_i | u_{\vn k m} \rangle \langle u_{\vn k m} | \hbar v_j(\vn k) | u_{\vn k n}\rangle \right] \\
\nonumber&\qquad\times\bigg\{ \frac{\mathcal E_{\vn k n}+\mathcal E_{\vn km} - 2\mathcal E_\mathrm{F}}{(\mathcal E_{\vn kn}-\mathcal E_{\vn km})^2} \mathrm{Im}\ln \frac{\mathcal E_{\vn km}-\mathcal E_\mathrm{F}-i\Gamma}{\mathcal E_{\vn kn}-\mathcal E_\mathrm{F}-i\Gamma} \\
&\qquad-\frac{2\Gamma}{(\mathcal E_{\vn kn}-\mathcal E_{\vn km})^2}\mathrm{Re}\ln \frac{\mathcal E_{\vn km}-\mathcal E_\mathrm{F}-i\Gamma}{\mathcal E_{\vn kn}-\mathcal E_\mathrm{F}-i\Gamma} \bigg\} \, .
\label{eq:dmi_gamma}
\end{align}
Note that the sums are not restricted to the occupied manifold but have to be performed in principle over all electronic states. In the clean limit of $\Gamma\rightarrow 0^+$, the second term in Eq.~\eqref{eq:dmi_gamma} vanishes such that the Berry phase expression~\eqref{eq:spiralization_zeroT} can be recovered. Since $\langle u_{\vn kn} | \mathcal T_i | u_{\vn k m} \rangle\rightarrow (\langle u_{\vn kn} | \mathcal T_i | u_{\vn k m} \rangle)^*$ and $\langle u_{\vn k m} | v_j(\vn k) | u_{\vn k n}\rangle\rightarrow -(\langle u_{\vn k m} | v_j(\vn k) | u_{\vn k n}\rangle)^*$ under time inversion, the DMI spiralization is apparently even with respect to magnetization reversal.

\paragraph{Relation to spin-orbit torques}Remarkably, the modern theory establishes an intimate connection between DMI and antidamping spin-orbit torques that are exerted on the magnetization due to the interplay of the spin-orbit interaction and an applied electric field $\vn E$. As we shall see below, these torques are in fact linked with the antisymmetric exchange interaction just like the anomalous Hall effect is related to the orbital magnetization in periodic solids~\cite{Thonhauser2011,Freimuth2014,Hanke2016}. The antidamping spin-orbit torques are ascribed to the non-trivial geometry of the mixed phase space of $\vn k$ and $\hat{\vn m}$ in terms of the so-called mixed Berry curvature $\Omega_{ij}^{\hat{\vn m}\vn k}$ of all occupied states~\cite{Freimuth2014}:
\begin{equation}
\Omega_{ij}^{\hat{\vn m}\vn k}(\vn k) =-2\hat{\vn e}_i\cdot \mathrm{Im}\sum_n^{\mathrm{occ}} \bigg[ \hat{\vn m}\times \bigg\langle \frac{ \partial u_{\vn kn}}{\partial \hat{\vn m}} \bigg| \frac{\partial u_{\vn k n}}{\partial k_j}\bigg\rangle \bigg] \, .
\label{eq:mixed_berry}
\end{equation}
In linear response to the electric field $\vn E$, the antidamping torque $\vn T= \tau \vn E$ is characterized by the torkance tensor $\tau$, which is determined by the sum of the mixed Berry curvature~\eqref{eq:mixed_berry} over the Brillouin zone~\cite{Freimuth2014}:
\begin{equation}
\tau_{ij} =-\frac{e}{N}\sum_{\vn k}\sum_n^{\mathrm{occ}} B_{ij}^n(\vn k)=-\frac{e}{N}\sum_{\vn k} \Omega_{ij}^{\hat{\vn m}\vn k}(\vn k) \, ,
\label{eq:torkance}
\end{equation}
where $e>0$ is the elementary positive charge.

\section{Method}

\subsection{Higher-dimensional Wannier functions}
Derivatives of the wave functions with respect to both crystal momentum and magnetization direction need to be evaluated in order to calculate from first principles the spiralization $D_{ij}$ and the torkance $\tau_{ij}$ according to their Berry phase expressions~\eqref{eq:spiralization_zeroT} and~\eqref{eq:torkance}, respectively. This has motivated us to develop an efficient generalization of Wannier functions~\cite{Hanke2015} for higher phase-space dimensions by performing Fourier transformations not only with respect to $\vn k$ but also with respect to additional parameters $\gpara$ (for example, the angles $\varphi$ and $\theta$) entering the Hamiltonian. In the following, we briefly review the main aspects of these higher-dimensional Wannier functions (HDWFs). Detailed discussions can be found in Ref.~\citen{Hanke2015}.

\paragraph{Orthogonality problem}Let us assume that the Hamiltonian $\hat H$ depends on an additional abstract parameter $\gpara$. Then, the eigenstates are Bloch states $|\psi_{\vn k \gpara n}\rangle = e^{i \vn k \cdot \vn r} |u_{\vn k \gpara n}\rangle$ that carry explicitly a dependence on this parameter, i.e., $\hat H |\psi_{\vn k \gpara n}\rangle = \mathcal{E}_{\vn k \gpara n} |\psi_{\vn k \gpara n}\rangle$, where the $\mathcal{E}_{\vn k \gpara n}$ denote the band energies. However, the eigenstates at $\gpara$ and $\gpara^\prime$ are not necessarily orthogonal since the Hamiltonians at different values of $\gpara$ are in general completely independent. This orthogonality problem obstructs using directly Fourier transformations of the usual Bloch states with respect to $\vn k$ and $\gpara$ to construct HDWFs~\cite{Hanke2015}. To naturally generalize Wannier functions to the multi-parameter case, we have to first restore the orthogonality by introducing an auxiliary space $\gpos$ as the reciprocal to the parameter space of $\gpara$. Thus, instead of taking the usual Bloch states, we consider in the construction of HDWFs orthogonal states $|\Phi_{\vn k \gpara n}\rangle$ in the higher-dimensional real space $(\vn r, \gpos)$. We define such states as the products of the physical Bloch states and an auxiliary orbital $|\artpsi_{\gpara}\rangle$:
\begin{equation}
 \Phi_{\vn k \gpara n}(\vn r,\gpos) = \psi_{\vn k \gpara n}(\vn r) \artpsi_{\gpara}(\gpos) \, .
\label{eq:gen_WF_prod}
\end{equation}
The required orthogonality of the product states $|\Phi_{\vn k \gpara n}\rangle$ is enforced by imposing the condition $\langle \artpsi_{\gpara} | \artpsi_{\gpara^\prime}\rangle = N_{\gpara}\delta_{\gpara\gpara^\prime}$, where $N_{\gpara}$ denotes the number of discrete $\gpara$-points. One particularly useful choice to satisfy this condition is the Bloch shape $|\artpsi_{\gpara}\rangle = e^{i\gpara\cdot\gpos} |\artu_{\gpara}\rangle$, where $|\artu_{\gpara}\rangle$ is lattice periodic in $\gpos$. As a consequence, the full product state~\eqref{eq:gen_WF_prod} is an eigenstate of the Hamiltonian $H$ with the eigenvalue $\mathcal E_{\vn k \gpara n}$, and assumes the desired Bloch-like form $|\Phi_{\vn k \gpara n}\rangle = e^{i\vn k \cdot \vn r}e^{i\gpara \cdot \gpos}|\phi_{\vn k \gpara n}\rangle$ with the lattice-periodic function $|\phi_{\vn k \gpara n}\rangle = |u_{\vn k \gpara n}\rangle \otimes |\artu_{\gpara}\rangle$.

\paragraph{Localization in real space}Discrete Fourier transformations of the orthogonal product states with respect to $\vn k$ and $\gpara$ can then be employed to define HDWFs as a meaningful generalization of Wannier functions~\cite{Hanke2015}:
\begin{equation}
 \gwann_{\vn R \gdlatt n}(\vn r,\gpos) = \frac{1}{N N_{\gpara}} \sum\limits_{\vn k \gpara m}  e^{-i \vn k\cdot \vn R} e^{-i \gpara \cdot \gdlatt} \gumat_{mn}^{(\vn k,\gpara)} \Phi_{\vn k \gpara m}(\vn r, \gpos) \, ,
\label{eq:HDWF_definition}
\end{equation}
which are labeled by the orbital index $n$, the direct lattice vector $\vn R$, and the additional lattice vector $\gdlatt$. The latter is conjugate to $\gpara$ like the direct lattice vector $\vn R$ is conjugate to the crystal momentum $\vn k$. Aiming at a maximal localization of these HDWFs, we determine uniquely the unitary gauge transformations $\gumat^{(\vn k,\gpara)}$ in Eq.~\eqref{eq:HDWF_definition} through minimization of the real-space spread $\Omega$ of \gwfshort{} that is given by the extension of the Marzari-Vanderbilt functional~\cite{Marzari1997,Hanke2015}:
\begin{equation}
 \Omega = \sum\limits_n \left(\langle \gwann_{\vn 0 \vn 0 n}|\mathfrak{r}^2|\gwann_{\vn 0 \vn 0 n}\rangle - \langle \gwann_{\vn 0 \vn 0 n}|\vn{\mathfrak{r}}|\gwann_{\vn 0 \vn 0 n}\rangle^2\right)\, .
\label{eq:HDWF_spread}
\end{equation}
Here, we introduced the higher-dimensional position operator $\vn{\mathfrak r}=(\vn r,\gpos)^{\text T}$ combining the position vectors $\vn r$ and $\gpos$.

In order to minimize systematically the spread~\eqref{eq:HDWF_spread}, we have extended the \wprog{} program~\cite{Mostofi2014}, which expects the following two matrices as a basic first-principles input. First, the overlaps $\langle \phi_{\vn k \gpara m} | \phi_{\vn k+\vn b_{\vn k} \, \gpara + \vn b_{\gpara} n}\rangle$ of the periodic parts of the product states at neighboring points in the $(\vn k, \gpara)$-space serve to calculate centers and spreads of HDWFs. Second, the projections $\langle \Phi_{\vn k \gpara m} | p_n \rangle$ onto localized trial orbitals $|p_n\rangle$ mark the starting point for the iterative spread minimization. Using our adapted version of \wprog{}, we are able to construct a single set of maximally-localized HDWFs from the electronic structure given on a coarse uniform grid of $(\vn k,\gpara)$-points~\cite{Hanke2015}.

\subsection{Higher-dimensional Wannier interpolation}

\paragraph{Multi-parameter Hamiltonian}After generating a single set of HDWFs via the modified maximal-localization procedure, we compute the periodic parts of Bloch-like functions:
\begin{equation}
 |\phi^{(\mathrm W)}_{\vn \kappa n}\rangle = \sum\limits_{\vn{\mathfrak{R}}} e^{-i\vn\kappa \cdot(\vn{\mathfrak{r}}-\vn{\mathfrak{R}})} |W_{\vn{\mathfrak{R}}n}\rangle = |u^{(\mathrm W)}_{\vn \kappa n}\rangle \otimes |\artu_{\gpara}^{\vphantom{(W)}}\rangle \, ,
 \label{eq:bloch_like_wannier}
\end{equation}
which are said to belong to the Wannier gauge ($\mathrm W$). In order to imitate for clarity the well-known terminology of the usual Wannier functions, we adopt here a simplified notation by introducing $\vn \kappa = (\vn k,\gpara)^{\mathrm T}$ as combined higher-dimensional vector of the crystal momentum and the additional parameter, and similarly $\vn{\mathfrak{R}}=(\vn R,\gdlatt)^{\mathrm T}$ for the direct lattice vectors. Because of the short-ranged hoppings $\langle \gwann_{\vn 0 n} | \hat H | \gwann_{\vn{\mathfrak{R}}m}\rangle$ between HDWFs,
\begin{equation}
 H^{(\mathrm W)}_{nm}(\vn \kappa) =\langle \phi^{(\mathrm W)}_{\vn \kappa n} | H_{\vn k} | \phi^{(\mathrm W)}_{\vn \kappa m}\rangle =\sum\limits_{\vn{\mathfrak{R}}} e^{i \vn \kappa\cdot \vn{\mathfrak R}} \langle \gwann_{\vn 0 n} | \hat H | \gwann_{\vn{\mathfrak R}m}\rangle
\label{eq:ham_interpol2}
\end{equation}
defines an efficient interpolation of the Hamiltonian matrix at any desired point $\vn \kappa$, even if this point is not contained in the coarse grid used for the construction of the HDWFs. Applying to Eq.~\eqref{eq:bloch_like_wannier} the unitary matrix $V(\vn\kappa)$ which diagonalizes the interpolated Hamiltonian~\eqref{eq:ham_interpol2} at every point $\vn \kappa$, we find the eigenstates $|\phi^{(\mathrm H)}_{\vn \kappa n}\rangle$ of this Hamiltonian as
\begin{equation}
 |\phi^{(\mathrm H)}_{\vn \kappa n}\rangle = \sum_m |\phi^{(\mathrm W)}_{\vn \kappa m}\rangle V^{\vphantom{(\mathrm W)}}_{mn}(\vn \kappa) = |u^{(\mathrm H)}_{\vn \kappa n}\rangle \otimes |\artu^{\vphantom{(\mathrm H)}}_{\gpara}\rangle \,.
 \label{eq:bloch_like_hamiltonian}
\end{equation}
Here, $|\phi^{(\mathrm H)}_{\vn \kappa n}\rangle$ and $|u^{(\mathrm H)}_{\vn \kappa n}\rangle = \sum_m |u^{(\mathrm W)}_{\vn \kappa m}\rangle V^{\vphantom{(\mathrm W)}}_{mn}(\vn \kappa)$ are said to belong to the Hamiltonian gauge ($\mathrm H$), in which the Hamiltonian is diagonal such that
\begin{equation}
 \langle \phi^{(\mathrm H)}_{\vn \kappa n} | H_{\vn k} | \phi^{(\mathrm H)}_{\vn \kappa m}\rangle =  \langle u^{(\mathrm H)}_{\vn \kappa n} | H_{\vn k} | u^{(\mathrm H)}_{\vn \kappa m}\rangle = \mathcal E_{\vn \kappa n} \delta_{nm} \, .
\end{equation}

\paragraph{Berry curvatures}We discuss now the generalized Wannier interpolation of pure and mixed Berry curvatures in the abstract phase space spanned by $\vn k$ and $\gpara$. The proposed scheme will be analogous to the case of conventional Wannier functions in Ref.~\citen{Wang2006} in that all quantities are finally evaluated within the Hamiltonian gauge. However, while the scheme in Ref.~\citen{Wang2006} is restricted to the Berry curvature in momentum space, we are able to study by means of the higher-dimensional interpolation additional curvatures that drive intriguing geometric phenomena such as the antidamping spin-orbit torques given by Eq.~\eqref{eq:torkance}. We start from the well-known expression of the Berry curvature matrix
\begin{equation}
\begin{split}
\Omega^{(\mathrm H)}_{nm,\alpha\beta}(\vn \kappa) &= \partial_{\alpha} A^{(\mathrm H)}_{nm,\beta}(\vn \kappa) - \partial_{\beta} A^{(\mathrm H)}_{nm,\alpha}(\vn \kappa) \\ &=-2\mathrm{Im}\langle \partial_\alpha u_{\vn \kappa n}^{(\mathrm H)} | \partial_\beta u_{\vn \kappa m}^{(\mathrm H)}\rangle\, ,
\end{split}
\label{eq:berry_curvature2}
\end{equation}
where $A^{(\mathrm H)}_{nm,\alpha}(\vn \kappa)=i\langle u^{(\mathrm H)}_{\vn \kappa n}| \partial_\alpha u^{(\mathrm H)}_{\vn \kappa m}\rangle$ is the Berry connection matrix, and $\partial_\alpha=\partial/\partial\kappa_\alpha$ with $\kappa_\alpha$ as $\alpha$th entry of $\vn \kappa$. The total Berry curvature of all occupied states is given by
\begin{equation}
 \Omega_{\alpha\beta}(\vn \kappa) = \sum_n f_n^{(\mathrm H)} \Omega^{(\mathrm H)}_{nn,\alpha\beta}(\vn \kappa) \, ,
 \label{eq:berry_curvature}
\end{equation}
with the occupation numbers $f_n^{(\mathrm H)}$. On the one hand, pure Berry curvatures are obtained if $\partial_\alpha$ and $\partial_\beta$ refer to the same derivative type. We may recover from Eq.~\eqref{eq:berry_curvature}, for example, the pure momentum Berry curvature that gives rise to the anomalous Hall conductivity. On the other hand, if $\alpha$ and $\beta$ refer to distinct variable types, we obtain mixed Berry curvatures, which determine magneto-electric coupling effects and the DMI in clean systems \cite{Freimuth2014,Kurebayashi2014,Hanke2017} if the abstract parameter $\gpara$ plays the role of the magnetization direction (cf. Eqs.~\eqref{eq:spiralization_zeroT} and~\eqref{eq:torkance}).

The Berry curvature~\eqref{eq:berry_curvature} can be accessed by taking the derivative of the wave functions $|u^{(\mathrm H)}_{\vn \kappa n}\rangle$ defined through Eq.~\eqref{eq:bloch_like_hamiltonian}:
\begin{equation}
 \partial_{\alpha} |u^{(\mathrm H)}_{\vn \kappa n}\rangle = \sum_m \partial_\alpha |u^{(\mathrm W)}_{\vn \kappa m}\rangle V_{mn}(\vn \kappa) + \sum_m |u^{(\mathrm W)}_{\vn \kappa m}\rangle \partial_\alpha V_{mn}(\vn \kappa) \, .
 \label{eq:diff_wavefunction1}
\end{equation}
Introducing for every matrix object $\mathcal O$ the abbreviation $\bar{\mathcal O}^{(\mathrm H)} = V^\dag \mathcal O^{(\mathrm W)} V$, we can rephrase the second term above since $\partial_\alpha V(\vn \kappa) = V(\vn \kappa) D_\alpha^{(\mathrm H)}(\vn \kappa)$, where
\begin{equation}
D_{mn,\alpha}^{(\mathrm H)}(\vn \kappa) = \begin{dcases} \frac{\bar H_{mn,\alpha}^{(\mathrm H)}}{\mathcal E_{\vn \kappa n}-\mathcal E_{\vn \kappa m}} &\text{, } n\neq m \\ 0 &\text{, } n=m \end{dcases} \, ,
\label{eq:Dalpha}
\end{equation}
which follows from standard pertubation theory in $\alpha$. In the previous definition $\bar H_\alpha^{(\mathrm H)} = V^\dag \partial_\alpha H^{(\mathrm W)} V$ with $H^{(\mathrm W)}$ given by Eq.~\eqref{eq:ham_interpol2}. Consequently, Eq.~\eqref{eq:diff_wavefunction1} assumes the form
\begin{equation}
 \partial_{\alpha} |u^{(\mathrm H)}_{\vn \kappa n}\rangle = \sum_m \partial_\alpha |u^{(\mathrm W)}_{\vn \kappa m}\rangle V_{mn}(\vn \kappa) + \sum_{m} |u^{(\mathrm H)}_{\vn \kappa m}\rangle D_{mn,\alpha}^{(\mathrm H)}(\vn \kappa) \, .
 \label{eq:diff_wavefunction2}
\end{equation}
Due to the adopted notation, the wave function derivative is formally identical to Eq.~(26) of Ref.~\citen{Wang2006} providing crystal momentum derivatives. However, we emphasize that the above equation is generalized to the higher-dimensional case, and thus allows us to extract wave function derivatives with respect to both $\vn k$ and $\gpara$. Based on Eq.~\eqref{eq:diff_wavefunction2}, we may carry out analogous algebra as in Ref.~\citen{Wang2006} using the identity
\begin{equation}
 A^{(\mathrm H)}_{nm,\alpha}(\vn \kappa) = \bar A^{(\mathrm H)}_{nm,\alpha}(\vn \kappa) + i D_{nm,\alpha}^{(\mathrm H)}(\vn \kappa)
 \label{eq:A_D}
\end{equation}
to arrive at the final expression for the Berry curvature:
\begin{equation}
\begin{split}
\Omega_{\alpha\beta}&= \sum_n f_n^{(\mathrm H)} \bar{\Omega}^{(\mathrm H)}_{nn,\alpha\beta} + i\sum_{nm}(f_m^{(\mathrm H)}-f_n^{(\mathrm H)})D_{nm,\alpha}^{(\mathrm H)} D_{mn,\beta}^{(\mathrm H)}\\
&+\sum_{nm}(f_m^{(\mathrm H)}-f_n^{(\mathrm H)}) \Big[D_{nm,\alpha}^{(\mathrm H)}\bar A^{(\mathrm H)}_{mn,\beta} - D_{nm,\beta}^{(\mathrm H)}\bar A^{(\mathrm H)}_{mn,\alpha}\Big] \, ,
\end{split}
\label{eq:berry_curvature_final}
\end{equation}
which provides access to pure or mixed Berry curvatures, depending on the explicit physical nature of the variables that are associated with the indices $\alpha$ and $\beta$.

To evaluate by means of generalized Wannier interpolation $\bar{\Omega}^{(\mathrm H)}_{\alpha\beta}(\vn \kappa)$, $\bar A^{(\mathrm H)}_{\alpha}(\vn \kappa)$, and $D_\alpha^{(\mathrm H)}(\vn \kappa)$, the corresponding quantities are needed first in the Wannier gauge. From Eq.~\eqref{eq:bloch_like_wannier}, Eq.~\eqref{eq:ham_interpol2}, and the choice $\langle \artu_{\gpara}|\partial_\alpha \artu_{\gpara}\rangle=0$ it follows that
\begin{align}
H^{(\mathrm W)}_{nm,\alpha}(\vn \kappa) &= i\sum_{\vn{\mathfrak R}} \mathfrak{R}_\alpha e^{i\vn \kappa \cdot \vn{\mathfrak R}} \langle \gwann_{\vn 0n} | \hat H | \gwann_{\vn{\mathfrak R} m}\rangle \, , \label{eq:ham_wannier_diff}\\
A^{(\mathrm W)}_{nm,\alpha}(\vn \kappa)  &= \sum_{\vn{\mathfrak R}} e^{i\vn \kappa \cdot \vn{\mathfrak R}} \langle \gwann_{\vn 0n} | \mathfrak{r}_\alpha | \gwann_{\vn{\mathfrak R} m}\rangle \, , \label{eq:A_wannier}\\
\begin{split}
\Omega^{(\mathrm W)}_{nm,\alpha\beta}(\vn \kappa) &= i\sum_{\vn{\mathfrak R}} e^{i\vn \kappa \cdot \vn{\mathfrak R}} ( \mathfrak{R}_\alpha \langle \gwann_{\vn 0n} | \mathfrak{r}_\beta | \gwann_{\vn{\mathfrak R} m}\rangle \\
&\qquad\qquad- \mathfrak{R}_\beta \langle \gwann_{\vn 0n} | \mathfrak{r}_\alpha | \gwann_{\vn{\mathfrak R} m}\rangle) \, . \label{eq:Omega_wannier}
\end{split}
\end{align}
Clearly, the hoppings and the matrix elements of the generalized position operator in the HDWF basis are necessary to calculate the desired quantities. These matrix elements are obtained by inverting Eq.~\eqref{eq:ham_interpol2} on the {\itshape ab initio} mesh. This yields for the hoppings
\begin{align}
\langle \gwann_{\vn 0 n}|\hat H|\gwann_{\vn{\mathfrak R} m}\rangle = \frac{1}{N_{\vn\kappa}} \sum\limits_{\vn\kappa} e^{-i\vn \kappa \cdot \vn{\mathfrak R}} \left[\gumat^{(\vn\kappa) \dag} \mathcal E_{\vn \kappa}\, \gumat^{(\vn \kappa)}\right]_{nm} \, ,
\label{eq:hoppingsW}
\end{align}
where $\mathcal E_{\vn \kappa}$ is a diagonal matrix containing the {\itshape ab initio} band energies, $N_{\vn \kappa}$ stands for the number of $\vn \kappa$-points on the coarse grid, and $\gumat^{(\vn \kappa)}$ is the maximal-localization gauge of Eq.~\eqref{eq:HDWF_definition}. Likewise, the higher-dimensional position matrix is found from inverting Eq.~\eqref{eq:A_wannier} on the coarse mesh:
\begin{equation}
\langle \gwann_{\vn 0 n} | \mathfrak{r}_\alpha | \gwann_{\vn{\mathfrak R} m}\rangle = \frac{i}{N_{\vn \kappa}} \sum\limits_{\vn \kappa} e^{-i \vn \kappa \cdot \vn{\mathfrak R}} \langle \phi^{(\mathrm W)}_{\vn \kappa n} | \partial_\alpha \phi^{(\mathrm W)}_{\vn \kappa m}\rangle \, .
\label{eq:posW}
\end{equation}
Using for the wave-function derivative in Eq.~\eqref{eq:posW} the finite-difference formula~\cite{Marzari1997}
\begin{equation}
|\partial_\alpha \phi^{(\mathrm{W})}_{\vn \kappa m}\rangle = \sum_{\vn b} w_b b_\alpha |\phi^{(\mathrm{W})}_{\vn \kappa+\vn bm}\rangle + O(b^2)\, ,
\label{eq:finitedifference}
\end{equation}
where $\vn b$ connects neighboring $\vn \kappa$-points and $w_b$ are appropriately chosen weights, we realize that the higher-dimensional positions~\eqref{eq:posW} are determined by the overlaps at neighboring grid points $\vn \kappa$ and $\vn \kappa + \vn b$, i.e., $\langle \phi_{\vn \kappa\vphantom{+\vn b} n}^{(\mathrm W)} | \phi_{\vn \kappa+\vn b m}^{(\mathrm W)}\rangle$ that enter already in the construction of maximally-localized HDWFs.

\paragraph{DMI spiralization}To elucidate the higher-dimensional Wannier interpolation of the spiralization $D_{ij}$, we exploit the analogies with the orbital magnetization noted earlier. A gauge-invariant formulation of the Wannier interpolation of orbital magnetization at zero temperature exists~\cite{Lopez2012}, and we extend this scheme here to evaluate Eq.~\eqref{eq:spiralization_zeroT} based on HDWFs. In the beginning, we rewrite the spiralization $D_{ij}$ as the sum of a local circulation $D_{ij}^\mathrm{lc}$ and an itinerant one $D_{ij}^\mathrm{ic}$ that read
\begin{align}
D_{ij}^\mathrm{lc} &= \frac{\hat{\vn e}_i}{NV}\cdot \mathrm{Im}\sum_{\vn k n}^{\mathrm{occ}}\hat{\vn m}\times \bigg\langle \frac{\partial u_{\vn \kappa n}}{\partial \hat{\vn m}} \bigg| H_{\vn k} - \mathcal E_\mathrm{F} \bigg| \frac{\partial u_{\vn \kappa n}}{\partial k_j}\bigg\rangle,
\label{eq:spiralization_local}\\
D_{ij}^\mathrm{ic} &= \frac{\hat{\vn e}_i}{NV}\cdot \mathrm{Im}\sum_{\vn k n}^{\mathrm{occ}}\hat{\vn m}\times \bigg\langle \frac{\partial u_{\vn \kappa n}}{\partial \hat{\vn m}} \bigg| \mathcal E_{\vn \kappa n} - \mathcal E_\mathrm{F} \bigg| \frac{\partial u_{\vn \kappa n}}{\partial k_j}\bigg\rangle  .
\label{eq:spiralization_itinerant}
\end{align}
Defining the $\vn \kappa$-dependent operator $\mathcal P=\sum_n^\mathrm{occ}|u_{\vn \kappa n}\rangle\langle u_{\vn \kappa n}|$ and its complement $\mathcal Q=1-\mathcal P$ that project onto the manifold of the occupied and unoccupied states, respectively, we introduce the ``geometric" objects
\begin{align}
{\mathcal F}_{\alpha\beta} &= \mathrm{Tr} \left[ (\partial_{\alpha} \mathcal P) \mathcal Q (\partial_{\beta} \mathcal P) \right] \, , \label{eq:FF}\\
{\mathcal G}_{\alpha\beta} &= \mathrm{Tr} \left[ (\partial_{\alpha} \mathcal P) \mathcal Q H_{\vn k} \mathcal Q (\partial_{\beta} \mathcal P) \right] \, , \label{eq:GG}\\
{\mathcal H}_{\alpha\beta} &= \mathrm{Tr} \left[ H_{\vn k} (\partial_{\alpha} \mathcal P) \mathcal Q (\partial_{\beta} \mathcal P) \right] \, , \label{eq:HH}
\end{align}
where dependence on $\vn \kappa=(\vn k,\hat{\vn m})^\mathrm{T}$ is implied, and $\mathrm{Tr}$ denotes the trace over the whole electronic Hilbert space. In resemblance to the orbital magnetization~\cite{Lopez2012}, the two parts~\eqref{eq:spiralization_local} and~\eqref{eq:spiralization_itinerant} of the spiralization can then be expressed as individually gauge-invariant properties of the ground state:
\begin{align}
D_{ij}^\mathrm{lc} &= \frac{\hat{\vn e}_i}{NV}\cdot \mathrm{Im}\sum_{\vn k n}^{\mathrm{occ}}\hat{\vn m}\times \left(\vn{\mathcal G}_{\hat{\vn m}k_j} - \mathcal E_{\mathrm F}\vn{\mathcal F}_{\hat{\vn m}k_j}\right)\, ,
\label{eq:spiralization_local_2}\\
D_{ij}^\mathrm{ic} &= \frac{\hat{\vn e}_i}{NV}\cdot \mathrm{Im}\sum_{\vn k n}^{\mathrm{occ}}\hat{\vn m}\times \left( \vn{\mathcal H}_{\hat{\vn m}k_j} - \mathcal E_{\mathrm F} \vn{\mathcal F}_{\hat{\vn m}k_j} \right)  \, ,
\label{eq:spiralization_itinerant_2}
\end{align}
where the short-hand vector notation $\vn{\mathcal F}_{\hat{\vn m}k_j}$, for example, stands for the geometric quantity of Eq.~\eqref{eq:FF} with $\partial_\alpha=\partial_{\hat{\vn m}}$ being the gradient with respect to the magnetization direction $\hat{\vn m}$ and $\partial_\beta = \partial_{k_j}$ as the derivative in the $j$th component of the crystal momentum. As a side remark, we note that the torkance~\eqref{eq:torkance} characterizing the antidamping spin-orbit torques assumes within this framework the alternative form
\begin{equation}
\tau_{ij}= \frac{2e}{N}\hat{\vn e}_i\cdot \mathrm{Im}\sum_{\vn kn}^{\mathrm{occ}} \hat{\vn m}\times \vn{\mathcal F}_{\hat{\vn m}k_j} \, .
\label{eq:torkance_2}
\end{equation}
By plugging the definition~\eqref{eq:FF} into this expression and using that $\mathrm{Im}\,\mathrm{Tr}\left[(\partial_{\alpha} \mathcal P)\mathcal P (\partial_{\beta}\mathcal P)\right]=0$, it is straightforward to verify that Eq.~\eqref{eq:torkance_2} is indeed equivalent to Eq.~\eqref{eq:torkance}. Similarly, we may prove the equivalence of the two different sets of expressions~\eqref{eq:spiralization_local}--\eqref{eq:spiralization_itinerant} and~\eqref{eq:spiralization_local_2}--\eqref{eq:spiralization_itinerant_2} for the local and itinerant circulations of the DMI spiralization tensor.

Although ${\mathcal F}_{\alpha\beta}$, ${\mathcal G}_{\alpha\beta}$, and ${\mathcal H}_{\alpha\beta}$ can in principle be evaluated under any gauge, we aim at calculating these phase-space properties in the Wannier gauge ($\mathrm{W}$), which turns out to be a convenient choice. To simplify the notation, we suppress from now on the superscript that was used before to indicate the Wannier gauge. The projection operator onto the occupied manifold spanned by the $M$ Bloch-like states, Eq.~\eqref{eq:bloch_like_wannier}, reads in this gauge $\mathcal P = \sum_{nm}^M|\phi_{\vn \kappa n}\rangle f_{nm} \langle \phi_{\vn \kappa m}|$, where the non-diagonal matrix $f$ is obtained by rotating the diagonal occupation matrix $f^{(\mathrm H)}$ into the Wannier gauge via $V^\dag(\vn \kappa)$. We define also $g=1-f$, where ``$1$" denotes the identity matrix in the considered $M$-dimensional subspace. In strong formal analogy to the derivations in Ref.~\citen{Lopez2012} for the orbital magnetization, we arrive at the following expressions for the phase-space objects that underlie the DMI spiralization:
\begin{align}
\nonumber\mathrm{Im}\, \mathcal F_{\alpha\beta} =& -\frac{1}{2}\mathrm{Re}\,\mathrm{tr}\left[f \Omega_{\alpha\beta}\right] + \mathrm{Im}\,\mathrm{tr}\left[ f A_\alpha g J_\beta + f J_\alpha g A_\beta \right] \\
&+\mathrm{Im}\,\mathrm{tr}\left[ f J_\alpha g J_\beta \right] \, , \\
\nonumber\mathrm{Im}\, \mathcal G_{\alpha\beta} =& -\frac{1}{2}\mathrm{Re}\,\mathrm{tr}\left[ f \Lambda_{\alpha\beta} \right] + \mathrm{Im}\,\mathrm{tr}\left[ f H f A_\alpha f A_\beta \right] \\*
&+\mathrm{Im}\,\mathrm{tr}\left[ f(J_\alpha g B_\beta - J_\beta g B_\alpha) + f J_\alpha g H g J_\beta \right] \, , \\
\nonumber\mathrm{Im}\, \mathcal H_{\alpha\beta} =&-\frac{1}{2}\mathrm{Re}\,\mathrm{tr}\left[ f H f \Omega_{\alpha\beta} \right] - \mathrm{Im}\,\mathrm{tr}\left[ f H f A_\alpha f A_\beta \right] \\
&+\mathrm{Im}\,\mathrm{tr}\left[ f H f ( A_\alpha g J_\beta + J_\alpha g A_\beta + J_\alpha g J_\beta) \right] \, ,
\end{align}
where all ingredients depend on $\vn \kappa$, and $\mathrm{tr}$ stands for the trace over the Bloch-like states, which we distinguish from the trace $\mathrm{Tr}$ in Eqs.~\eqref{eq:FF}--\eqref{eq:HH}. The matrices $H$, $A_\alpha$, and $\Omega_{\alpha\beta}$ in the Wannier gauge were already defined in Eqs.~\eqref{eq:ham_interpol2},~\eqref{eq:A_wannier}, and~\eqref{eq:Omega_wannier}, respectively, based on hopping and position elements of the HDWFs. Furthermore, we introduced in the above equations $J_\alpha = i V D_\alpha^{(\mathrm H)} V^\dagger$ with $D_\alpha^{(\mathrm H)}$ defined in Eq.~\eqref{eq:Dalpha} as well as the additional matrices
\begin{equation}
B_{nm,\alpha} = \sum_{\vn{\mathfrak R}} e^{i\vn \kappa \cdot \vn{\mathfrak R}} \langle \gwann_{\vn 0n} | \hat H \mathfrak{r}_\alpha^\prime | \gwann_{\vn{\mathfrak R} m}\rangle \, , \label{eq:B_wannier} 
\end{equation}
where $\vn{\mathfrak r}^\prime=\vn{\mathfrak r}-\vn{\mathfrak R}$, and $\Lambda_{\alpha\beta}=i(C_{\alpha\beta}-C_{\alpha\beta}^\dagger)$ with
\begin{equation}
C_{nm,\alpha\beta} = \sum_{\vn{\mathfrak R}} e^{i\vn \kappa \cdot \vn{\mathfrak R}} \langle \gwann_{\vn 0n} | \mathfrak{r}_\alpha \hat H \mathfrak{r}_\beta^\prime | \gwann_{\vn{\mathfrak R} m}\rangle \, . \label{eq:C_wannier} 
\end{equation}
In contrast to the higher-dimensional Wannier interpolation of Berry curvatures, matrix elements of $\hat H \vn{\mathfrak r}^\prime$ and $\vn{\mathfrak r} \hat H \vn{\mathfrak r}^\prime$ in the basis of HDWFs are required to evaluate the DMI spiralization tensor in periodic systems. Inverting the Fourier transformations in Eqs.~\eqref{eq:B_wannier} and~\eqref{eq:C_wannier} on the coarse $\vn \kappa$-grid, we arrive at explicit expressions for these matrix elements:
\begin{align}
\langle \gwann_{\vn 0 n} | \hat H \mathfrak{r}_\alpha^\prime | \gwann_{\vn{\mathfrak R} m}\rangle &= \frac{i}{N_{\vn \kappa}} \sum\limits_{\vn \kappa} e^{-i \vn \kappa \cdot \vn{\mathfrak R}} \langle \phi_{\vn \kappa n} |H_{\vn k} | \partial_\alpha \phi_{\vn \kappa m}\rangle \, , \label{eq:B_wannier_2} \\
\langle \gwann_{\vn 0 n} | \mathfrak r_\alpha \hat H \mathfrak{r}_\beta^\prime | \gwann_{\vn{\mathfrak R} m}\rangle &= \frac{1}{N_{\vn \kappa}} \sum\limits_{\vn \kappa} e^{-i \vn \kappa \cdot \vn{\mathfrak R}} \langle \partial_\alpha \phi_{\vn \kappa n} |H_{\vn k} | \partial_\beta \phi_{\vn \kappa m}\rangle \, .
\label{eq:C_wannier_2}
\end{align}
If we make use of the finite-difference formula~\eqref{eq:finitedifference} to express the derivatives of the wave functions, we realize that Eq.~\eqref{eq:B_wannier_2} is determined by the band energies and the overlaps $\langle \phi_{\vn \kappa\vphantom{+\vn b} n} | \phi_{\vn \kappa+\vn b m}\rangle$. However, the matrix elements of $\vn{\mathfrak r} \hat H \vn{\mathfrak r}^\prime$ given by Eq.~\eqref{eq:C_wannier_2} rely on new information encoded in overlaps $\langle \phi_{\vn \kappa + \vn b_1 n} | H_{\vn k} | \phi_{\vn \kappa +\vn b_2 m}\rangle$ of the Hamiltonian between states at neighboring points $\vn \kappa+\vn b_1$ and $\vn \kappa + \vn b_2$. Even though they do not enter directly in the construction of HDWFs, these quantities are relevant as they enable us to perform the higher-dimensional Wannier interpolation of the DMI spiralization. Therefore, we have implemented their computation from first principles in the \fleur{} code~\cite{fleur}.

\paragraph{Velocity and torque operators}Alternatively to the previously discussed methods, the torkance $\tau_{ij}$ and the spiralization $D_{ij}$ can be obtained by interpolating explicitly the velocity and torque operators, followed by calculating~\eqref{eq:spiralization_A} and~\eqref{eq:spiralization_B}. This amounts to performing the higher-dimensional Wannier interpolation of the matrix elements $\langle u^{(\mathrm H)}_{\vn \kappa n} | \partial_\alpha H_{\vn k} | u^{(\mathrm H)}_{\vn \kappa m} \rangle$, where $\partial_\alpha$ differentiates with respect to either the crystal momentum or the additional parameter $\gpara$, which grants access to the torque operator if $\gpara=\hat{\vn m}$. By virtue of the Hellmann-Feynman theorem and Eq.~\eqref{eq:A_D}, it follows that
\begin{equation}
 \langle u^{(\mathrm H)}_{\vn \kappa n} | \partial_\alpha H_{\vn k} | u^{(\mathrm H)}_{\vn \kappa m} \rangle = \bar H_{nm,\alpha}^{(\mathrm H)} - i(\mathcal{E}_{\vn \kappa m}-\mathcal{E}_{\vn \kappa n})\bar A_{nm,\alpha}^{(\mathrm H)} \, .
 \label{eq:vt_matrix_elements2}
\end{equation}
Based on the matrix elements~\eqref{eq:vt_matrix_elements2} of the generic adiabatic interaction $\partial_\alpha H_{\vn k}$, the interpolated velocity ($\partial_\alpha=\partial_{\vn k}$) assumes the form
\begin{equation}
 \hbar \vn v_{nm}(\vn \kappa) = \langle u_{\vn \kappa n}^{(\mathrm H)} | \partial_{\vn k} H_{\vn k} | u_{\vn \kappa m}^{(\mathrm H)}\rangle = \bar H_{nm,\vn k}^{(\mathrm H)} - i(\mathcal{E}_{\vn \kappa m}-\mathcal{E}_{\vn \kappa n})\bar A_{nm,\vn k}^{(\mathrm H)} \, ,
 \label{eq:velocity_interpolation}
\end{equation}
and analogously for the torque operator ($\partial_\alpha=\partial_{\hat{\vn m}}$):
\begin{equation}
\begin{split}
 \vn{\mathcal T}_{nm}(\vn \kappa) &= \left\langle u_{\vn \kappa n}^{(\mathrm H)} \bigg|\hat{\vn m} \times \frac{\partial H}{\partial \hat{\vn m}}\bigg| u_{\vn \kappa m}^{(\mathrm H)}\right\rangle \\
 &=\hat{\vn m} \times \left[ \bar H_{nm,\hat{\vn m}}^{(\mathrm H)} - i(\mathcal{E}_{\vn \kappa m}-\mathcal{E}_{\vn \kappa n})\bar A_{nm,\hat{\vn m}}^{(\mathrm H)} \right]
\label{eq:torque_interpolation}
\end{split}
\end{equation}
The higher-dimensional interpolation of Eqs.~\eqref{eq:velocity_interpolation} and~\eqref{eq:torque_interpolation} relies only on the hoppings~\eqref{eq:hoppingsW} and the positions~\eqref{eq:posW} in the basis of HDWFs but not on the matrix elements of $\hat H \vn{\mathfrak r}^\prime$ and $\vn{\mathfrak r} \hat H \vn{\mathfrak r}^\prime$. Nevertheless, we find that $D_{ij}$ obtained through this approach agrees well with the full scheme discussed before, which interpolates the DMI spiralization in close analogy to the modern theory of orbital magnetization.

\section{Results}

\begin{figure*}[t]
\centering
\includegraphics{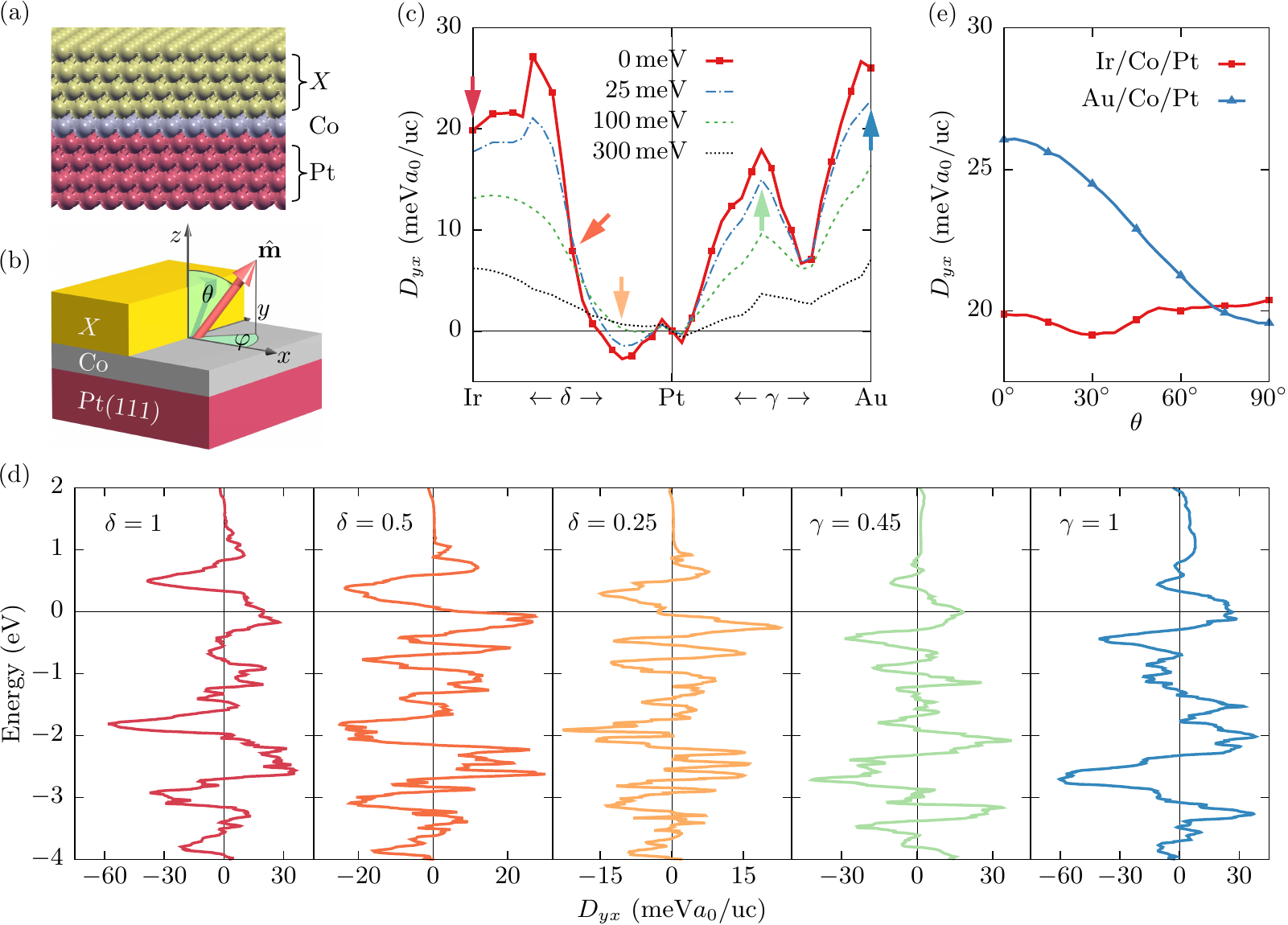}
\caption{(Color online) (a)~Crystal structure of the heterostructure, which comprises in total nine atomic layers, including the Pt(111) substrate, the ferromagnetic Co monolayer, and the overlayers $X$. (b)~The magnetization direction $\hat{\vn m}$ of the Co atoms is represented by the polar angle $\theta$ and the azimuthal angle $\varphi$. In this work, we consider the case of $\varphi=0$, i.e., the magnetization direction is specified by the single angle $\theta$. (c)~Dzyaloshinskii-Moriya spiralization $D_{yx}$ as a function of the overlayer composition in Ir$_\delta$Pt$_{1-\delta}$/Co/Pt and Au$_\gamma$Pt$_{1-\gamma}$/Co/Pt, respectively, where the magnetization is oriented perpendicular to the film plane. While the solid red curve corresponds to the clean limit, the thinner lines include the effect of disorder due to a finite band broadening $\Gamma$ of $25\,$meV, $100\,$meV, and $300\,$meV. The spiralization is given in the units meV$a_0$/uc, where $a_0$ is Bohr's radius and ``uc" stands for the in-plane unit cell. (d)~Energy dependence of $D_{yx}$ in the clean limit for the various alloy concentrations that are marked in (c) by arrows. The energy scale is relative to the corresponding Fermi level. (e)~Anisotropy of $D_{yx}$ with the polar angle $\theta$ in the inversion-asymmetric heterostructures Ir/Co/Pt and Au/Co/Pt.}
\label{fig:fig1}
\end{figure*}

\begin{figure*}[t]
\centering
\includegraphics{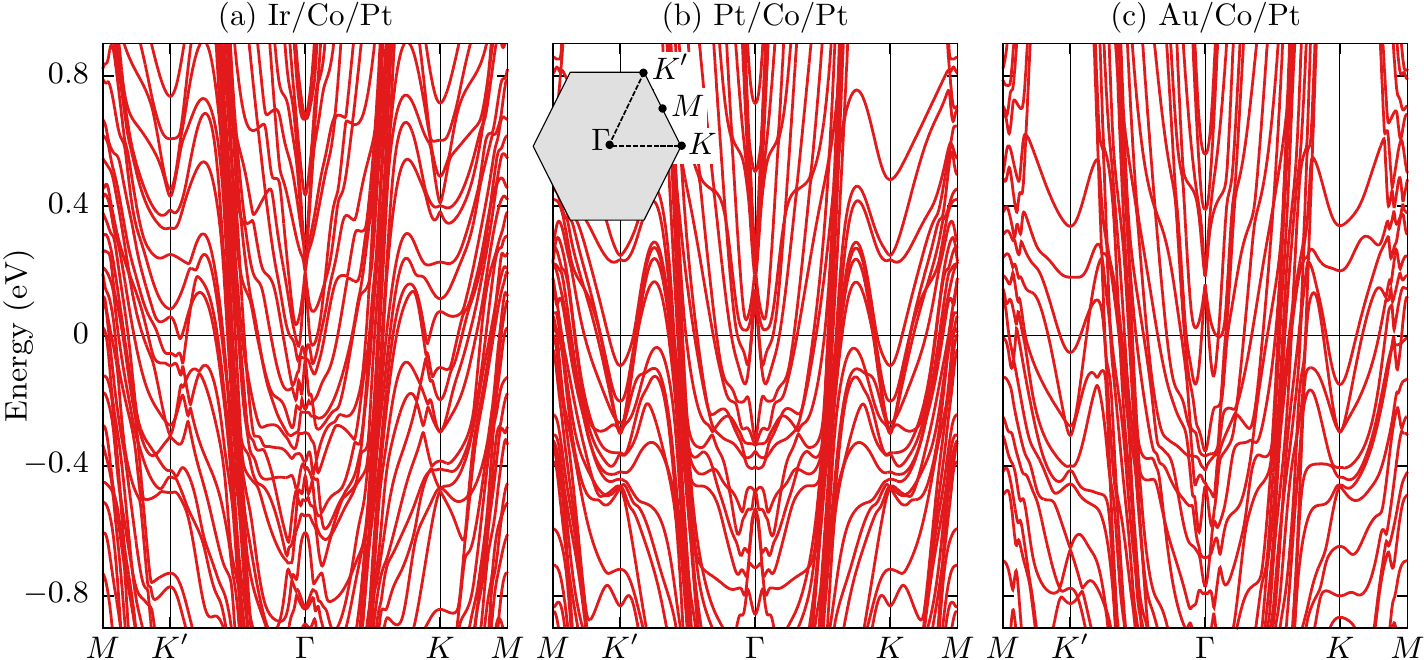}
\caption{(Color online) Electronic band structure along the high-symmetry lines of the two-dimensional hexagonal Brillouin zone (see inset of the middle panel) in the three stochiometric heterostructures of (a) Ir/Co/Pt, (b) Pt/Co/Pt, and (c) Au/Co/Pt. The energy scale is relative to the corresponding Fermi level. In all of these cases, the Co layer is magnetized perpendicular to the film plane, which corresponds to the polar angle $\theta=0$.}
\label{fig:fig2}
\end{figure*}

\paragraph{Computational details}We apply the Berry phase theory and the higher-dimensional Wannier interpolation to gain insights into the DMI spiralization in the Co-based heterostructures Ir$_\delta$Pt$_{1-\delta}$/Co/Pt and Au$_\gamma$Pt$_{1-\gamma}$/Co/Pt, which comprise in total nine atomic layers as shown in Fig.~\ref{fig:fig1}(a). Using the full-potential linearized augmented-plane-wave code \fleur{}~\cite{fleur}, we performed self-consistent density functional theory calculations of the electronic structure. Starting from the experimental parameters for the lattice structure of Pt/Co/Pt, we relaxed the interface layers and kept the resulting ionic positions fixed among all systems. Except for Co where the muffin-tin radius was set to $2.23\,a_0$ with $a_0$ as Bohr's radius, we chose the muffin-tin radii of all atoms as $2.29\,a_0$, and employed a plane-wave cutoff of $4.0\,a_0^{-1}$. While we treated exchange and correlation effects within the generalized gradient approximation, similar results can be obtained when using the local density approximation to density functional theory. To account for the effect of substitutional alloying of the top Pt layers with Ir and Au as controlled by $\delta$ and $\gamma$, respectively, the simple yet predictive virtual crystal approximation~\cite{Bellaiche2000} was used. Based on the calculated electronic structure on a coarse mesh of $8$$\times$$8$ $\vn k$-points and $8$ different angles $\theta$ that sample $[0,2\pi)$, we used our extension of the \wprog{} program~\cite{Mostofi2014,Hanke2015} to construct a single set of $162$ HDWFs out of $228$ Bloch bands, with the frozen window extending up to about $2\,$eV above the corresponding Fermi energy. The higher-dimensional Wannier interpolation was subsequently applied to Eqs.~\eqref{eq:spiralization_zeroT} and~\eqref{eq:dmi_gamma} in order to integrate efficiently over $1024$$\times$$1024$ $\vn k$-points in the full Brillouin zone.

\begin{figure*}[t]
\centering
\includegraphics{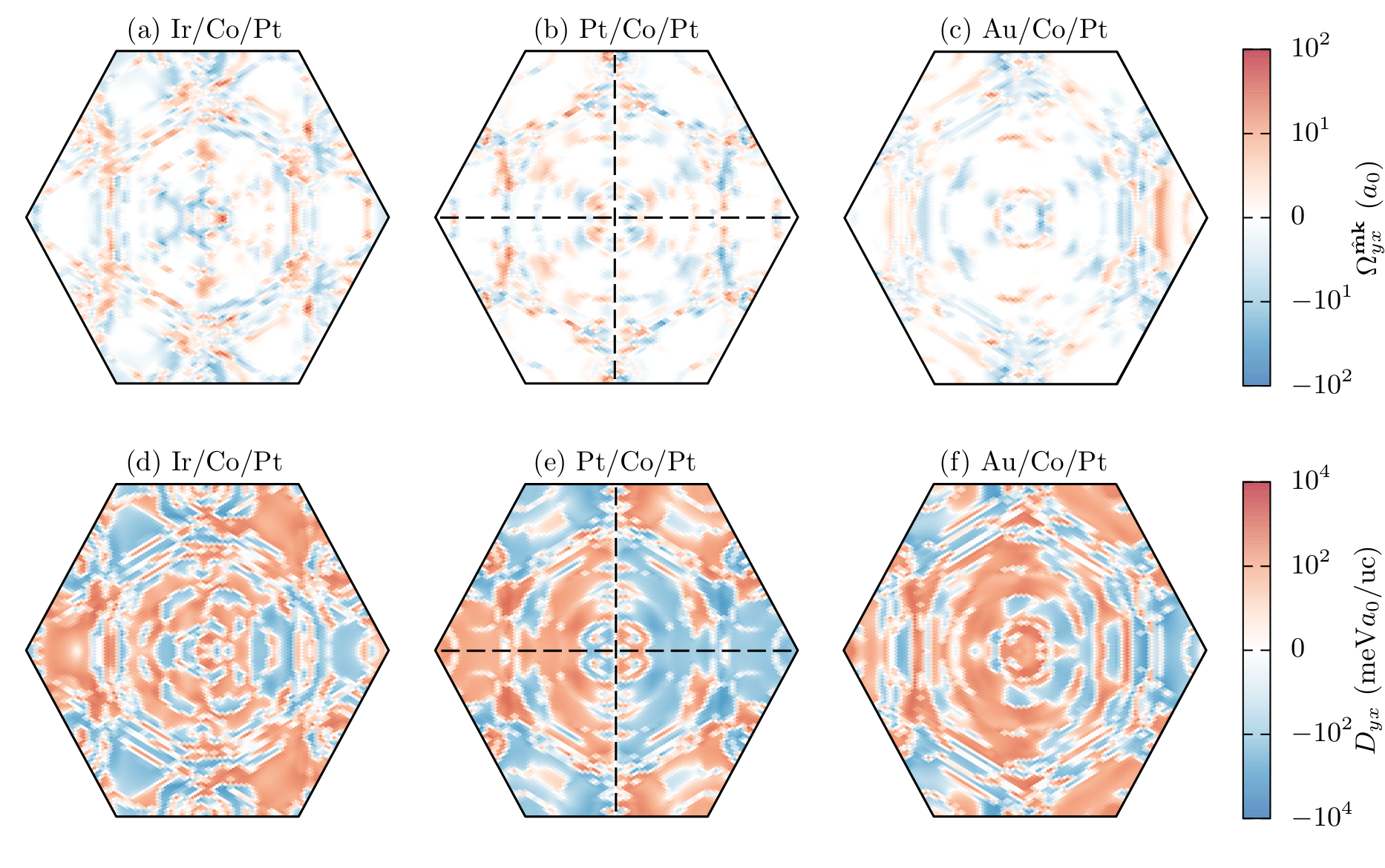}
\caption{(Color online) Distribution of the mixed Berry curvature $\Omega^{\hat{\vn m}\vn k}_{yx}$ (top) and the Dzyaloshinskii-Moriya spiralization $D_{yx}$ (bottom) of all occupied bands in the hexagonal Brillouin zone for the three stochiometric heterostructures Ir/Co/Pt, Pt/Co/Pt, and Au/Co/Pt. The Co monolayer in these thin films is perpendicularly magnetized. In the labels, $a_0$ denotes Bohr's radius and ``uc" stands for the in-plane unit cell. Note the logarithmic color scale.}
\label{fig:fig3}
\end{figure*}

\paragraph{Dependence on the overlayer}Figure~\ref{fig:fig2} depicts the complex electronic structure of the perpendicularly magnetized heterostructures Ir/Co/Pt, Pt/Co/Pt, and Au/Co/Pt around the Fermi level along high-symmetry lines in the hexagonal Brillouin zone. While the symmetric band structure of Pt/Co/Pt is the same at $\vn k$ and $-\vn k$, the band dispersions of both Ir/Co/Pt and Au/Co/Pt are different around the two valleys $K$ and $K^\prime$ as a consequence of the lack of spatial inversion symmetry. In the latter trilayers with out-of-plane magnetization, the DMI spiralization is described by an antisymmetric tensor due to the three-fold rotational symmetry around the axis normal to the film, which amounts to $D_{xx}=D_{yy}=0$ and $D_{yx}=-D_{xy}$. We present in Fig.~\ref{fig:fig1}(c) the variation of the only non-trivial entry $D_{yx}$ upon altering the composition ratio of the top layers by substitutional doping of Pt with Ir or Au. Remarkably, the DMI spiralization displays a complex non-monotonic dependence on the alloy composition, and even reverses its sign near $\delta=0.25$ that corresponds to moderate doping with Ir. In the regime of strong doping, for which either $\gamma\approx 1$ or $\delta\approx 1$, the systems with Au-rich overlayers exhibit larger values of $D_{yx}$ as compared to the Ir counterparts. Overall, our results for the Co-based heterostructures demonstrate that proper electronic-structure engineering, for example, via alloying provides an efficient tool to tailor magnitude and sign of the DMI in trilayer systems. Finally, we remark that including the effect of disorder by means of a constant broadening $\Gamma=25\,$meV of the energy bands smoothens the curve in Fig.~\ref{fig:fig1}(c) but hardly affects the qualitative behavior of the DMI spiralization.

In order to identify the fingerprints of the underlying electronic structure that manifest in the DMI spiralization, we display in Fig.~\ref{fig:fig1}(d) the variation of $D_{yx}$ with the position of the Fermi level for selected composition ratios. The computed oscillatory behavior is strongly reminiscent of the energy dependence of the anomalous Hall conductivity and the orbital magnetization in thin films~\cite{Hanke2016}. The peak structure in Fig.~\ref{fig:fig1}(d) can be ascribed to the positions of the $3d$ states in Co and the $5d$ states of the heavy metals as well as to their mutual hybridization. Since the $5d$ states of the spin-orbit active overlayers shift towards lower energies when moving from Ir to Au, also the major features of the DMI spiralization tend to be at lower energies. When these states hybridize with other orbitals, the intricate shape of the DMI curve is modified additionally. As the relevant $d$ states are not accessible at higher energies, the value of $D_{yx}$ is strongly suppressed above $1\,$eV.

By monitoring the momentum-resolved distribution of the spiralization over the Brillouin zone, we uncover in Fig.~\ref{fig:fig3} the microscopic origin of the DMI and compare it with the mixed Berry curvature $\Omega^{\hat{\vn m}\vn k}_{yx}$ of all occupied bands given by Eq.~\eqref{eq:mixed_berry}. These objects measure the torque-velocity correlation and as such naturally do not follow the rotational symmetries of the considered systems Ir/Co/Pt, Pt/Co/Pt, and Au/Co/Pt in momentum space since they differentiate only with respect to a single component of the crystal momentum. As a consequence, the calculated distributions of these quantities at $\vn k$ and $-\vn k$ are generally not at all linked by symmetry. However, owing to the preserved inversion symmetry, the case of Pt/Co/Pt marks an exception for which the DMI spiralization and the mixed Berry curvature at opposite crystal momenta are both exactly inverted, leading to a consistent cancellation and thus no net effect. In contrast to $\Omega^{\hat{\vn m}\vn k}_{yx}$ that is sharply peaked in narrow regions, the distribution of $D_{yx}$ displays a more rich texture with crucial background contributions arising from broad areas of the Brillouin zone. Remarkably, the total DMI spiralization is determined by the integral of large but strongly varying local contributions in momentum space, which renders electrical currents a promising means to promote drastic changes of the DMI spiralization mediated by the induced non-equilibrium population of the states.

\paragraph{Anisotropy with magnetization direction}A major benefit of the modern theory of DMI spiralization is the readily accessible dependence of this response tensor on the magnetic orientation of the ferromagnet. The higher-dimensional Wannier interpolation is designed to fully exploit this asset. Whereas the crystal symmetries dictate that the spiralization is an antisymmetric tensor for $\hat{\vn m}$ perpendicular to the film, this holds no longer if the magnetization points along a generic direction. In the case of a constant azimuthal angle $\varphi=0$, however, the element $D_{yx}$ remains by far the most dominant component in the studied Co-based trilayers. The dependence of $D_{yx}$ on the magnetization direction represented by the polar angle $\theta$ in Ir/Co/Pt and Au/Co/Pt is displayed in Fig.~\ref{fig:fig1}(e). Although the DMI spiralization varies with $\theta$ in both materials, the anisotropy is much more prominent for the Au/Co/Pt system as compared to the case of the Ir overlayers. In the former heterostructure, the original value of $26\,$meV$a_0$/uc for $D_{yx}$ is reduced considerably by about one quarter to $20\,$meV$a_0$/uc if the magnetization direction resides in the film plane instead of being oriented perpendicular to it.

\section{Summary}
We present an advanced first-principles technique to evaluate the Dzyaloshinskii-Moriya interaction (DMI) in its modern theory as well as Berry curvatures in the mixed space of crystal momentum and magnetization direction based on a higher-dimensional Wannier interpolation. This generic method can be used to study the Hamiltonian evolution under slowly varying parameters in various fields of physics. Here, we apply it to understand the microscopic origin of the DMI and its anisotropy in Co-based trilayers Ir$_\delta$Pt$_{1-\delta}$/Co/Pt and Au$_\gamma$Pt$_{1-\gamma}$/Co/Pt. Studying the dependence on the overlayer composition, we demonstrate that both magnitude and sign of the DMI in these ferromagnetic heterostructures can be tuned by means of electronic-structure engineering. In particular, we anticipate characteristic sign changes of the spiralization to manifest in systems with moderate Ir concentrations, which we proclaim as promising materials for experiments to investigate in more detail.

\section*{Acknowledgments}
\begin{acknowledgments}
We thank Bertrand Dup\'e, Christopher H. Marrows, and Bernd Zimmermann for many fruitful discussions. We gratefully acknowledge computing time on the supercomputers JUQUEEN and JURECA at J\"ulich Supercomputing Center as well as at the JARA-HPC cluster of RWTH Aachen, and funding from the German Research Foundation (DFG) under Grant No. MO 1731/5-1 as well as from the European Unions Horizon 2020 research and innovation programme under grant agreement number 665095 (FET-Open project MAGicSky).
\end{acknowledgments}

\end{document}